\documentclass[11pt]{article}
\usepackage{latexsym}  
\usepackage{amssymb}
\usepackage{graphicx}
\usepackage{amsmath}

\begin{document}

\hspace*{9cm} {OU-HET-741/2012; MISC-2012-05}

\begin{center}
{\Large\bf Family Gauge Bosons with an Inverted Mass Hierarchy}

\vspace{4mm}

\vspace{4mm}
{\bf Yoshio Koide$^{a,b}$ and Toshifumi Yamashita$^b$}

${}^{a}$ {\it Department of Physics, Osaka University, 
Toyonaka, Osaka 560-0043, Japan} \\
{\it E-mail address: koide@het.phys.sci.osaka-u.ac.jp}

${}^b$ {\it   MISC, 
Kyoto Sangyo University,  
Kyoto 603-8555, Japan}\\
{\it E-mail address: tyamashi@cc.kyoto-su.ac.jp}

\date{\today}
\end{center}

\vspace{3mm}
\begin{abstract}
A model that gives family gauge bosons with an inverted 
mass hierarchy is proposed, stimulated by Sumino's cancellation 
mechanism for the QED radiative correction to the charged 
lepton masses.
The Sumino mechanism cannot straightforwardly be applied to SUSY 
models because of the non-renormalization theorem.   
In this paper, an alternative model which is applicable 
to a SUSY model is proposed.
It is essential that family gauge boson masses $m(A_i^j)$ 
in this model is given by an inverted mass hierarchy 
$m(A_i^i) \propto 1/\sqrt{m_{ei}}$, in contrast to 
$m(A_i^i) \propto \sqrt{m_{ei}}$ in the original Sumino model.
Phenomenological meaning of the model is also investigated.  
In particular, we notice a deviation from the $e$-$\mu$ universality in the
tau decays.
\end{abstract}

\vspace{3mm}

\noindent{\large\bf 1 \ Introduction}

It seems to be meaningful to consider that the flavor physics can be understood on 
the basis of a family symmetry \cite{FamilySymm}. 
Regrettably, since a constraint from the observed $K^0$-$\bar{K}^0$ mixing is
very tight, the family gauge boson masses must be super heavy, so that 
it is hard to observe such gauge boson effects in terrestrial experiments. 
Recently, one positive effect of the existence of family gauge bosons has
been pointed out by Sumino \cite{Sumino09PLB,Sumino09JHEP}.
Since we propose a model with a Sumino-like mechanism in the present paper, 
we first give a brief review of the Sumino mechanism.
  
We know a miraculous formula for the charged lepton masses \cite{Koidemass}:
$$
K \equiv \frac{m_e +m_\mu + m_\tau}{
   (\sqrt{m_e} + \sqrt{m_\mu} + \sqrt{m_\tau})^2} 
   = \frac{2}{3} , 
\eqno(1.1)
$$
which is satisfied with the order 
of $10^{-5}$ for the pole masses, i.e. 
$K^{pole}=(2/3)\times (0.999989 \pm 0.000014)$ 
\cite{PDG10}. 
However, in conventional flavor models, ``masses" 
do not mean ``pole masses",  but ``running masses". 
The formula (1.1) is only valid with 
the order of $10^{-3}$ for the running masses, 
e.g. $K(\mu)=(2/3)\times (1.00189 \pm 0.00002)$ 
at $\mu =m_Z$. 
The deviation of $K(\mu)$ from $K^{pole}$ is 
caused by a logarithmic term $m_{ei}\log(\mu/m_{ei})$ 
in the QED radiative correction term \cite{Arason} 
$$
m_{ei}(\mu) = m_{ei}^{pole} \left[ 1-\frac{\alpha_{em}(\mu)}{\pi}
\left(1 +\frac{3}{4} \log \frac{\mu^2}{m_{ei}^2(\mu)} \right)
\right].
\eqno(1.2)
$$
Note that the value of $K$ is invariant under the transformation
$$
m_{ei} \rightarrow  m_{ei}(1+ \varepsilon_0 +\varepsilon_i) ,
\eqno(1.3)
$$
when $\varepsilon_i=0$, 
where 
$\varepsilon_0$ and $\varepsilon_i$ are factors which are 
independent of and dependent on the family-number $i$ ($i=1,2,3$),
respectively.
If the logarithmic term in the radiative correction (1.2) 
due to the photon can be cancelled by a some additional effect, 
the relation $K^{pole} = K(\Lambda)\equiv 2/3$ can be 
satisfied 
($\Lambda$ is an energy scale at which $K=2/3$ is given). 

Sumino \cite{Sumino09PLB} has seriously taken why the mass formula 
$K=2/3$ is so remarkably satisfied with the pole masses,  
and proposed a cancellation mechanism of the logarithmic term 
in the radiative correction (1.2).
He has assumed that the family symmetry is local, 
and that the logarithmic term 
is canceled by 
that due to the family gauge bosons.  
In the Sumino model, the left- and right-handed
charged leptons $e_{Li}$ and $e_{Ri}$ are assigned to ${\bf 3}$
and ${\bf 3}^*$ of a U(3) family symmetry, respectively. 
(A similar fermion assignment has been proposed by 
Appelquist, Bai and Piai \cite{Appelquist06}.)
The charged lepton mass term is 
generated by a would-be Yukawa interaction
$$
H_e = \frac{y_e}{\Lambda^2} 
\bar{\ell}_{L}^i \Phi^e_{i \alpha}
 \Phi^{eT}_{\alpha j} e_{R}^j H ,
\eqno(1.4)
$$
where $i$ and $\alpha$ are indices of U(3) and O(3), respectively,
$H$ is the Higgs scalar in the standard non-SUSY model, 
and $\ell_L =(\nu, e^-)$.
The VEV matrix $\langle \Phi_e \rangle$
is assumed as 
$$
\langle \Phi_e \rangle = {\rm diag}(v_1,v_2,v_3) 
\propto {\rm diag}(\sqrt{m_e}, \sqrt{m_\mu}, \sqrt{m_\tau}) .
\eqno(1.5)
$$
Then, the family gauge boson masses $m(A_i^j)$ are given by 
$$
m^2(A^i_j) \propto 
\langle \Phi^e_{i\alpha} \rangle \langle (\Phi^e)^{\dagger \alpha i}\rangle 
+\langle \Phi^e_{j\alpha} \rangle \langle (\Phi^e)^{\dagger \alpha j}\rangle
\propto m_{ei}+m_{ej} .
\eqno(1.6)
$$
Since in the Sumino model, the charged lepton fields $(e_L, e_R)$ are 
assigned to $(e_L, e_R) \sim ({\bf 3}, {\bf 3}^*)$ of the U(3), 
the family gauge bosons in the off-diagonal elements $A_i^j$ ($i\neq j$)
cannot contribute to the radiative corrections.  
Then, the cancellation takes place between $\log m_{ei}$ in the QED diagram 
and $\log m(A_i^i) \propto \log m_{ei}$ in the family gauge boson diagram.
(Of cause, the family gauge boson coupling constant $g_F$ must satisfy
a relation $g_F^2/4 \simeq e^2$.)
As a result, we can obtain $K(\Lambda) =K^{pole}$. 

In Ref.\cite{Sumino09PLB}, and also in this paper, 
it is assumed that the formula (1.1) is already 
given at an energy scale $\Lambda$, and is not discussed how to 
derive the formula\footnote{
The first attempt to understand the mass formula (1.1) from the
bilinear form of $\Phi_e$ has been done by assuming 
a U(3) family symmetry and a ``nonet" ansatz for $\Phi$ 
\cite{Koide90MPL}.  
For more plausible derivation of the formula, see 
Ref. \cite{Sumino09JHEP}, where the model is based on a U(9) 
family symmetry.  
} 
.

Now, it is interesting to apply this Sumino mechanism to a  
supersymmetric (SUSY) scenario. 
It should be noted that a vertex correction is, in 
general, vanishing in the SUSY scenario, so that the Sumino mechanism
cannot be applied to SUSY models 
straightforwardly. 
In this paper, we investigate how to restore the Sumino mechanism
in such models. 
The essential idea in the present model is as follows: 
The cancellation in the Sumino mechanism occurs   
due to the vertex correction diagram, 
while that in the present paper does due to the wave function 
renormalization diagrams.  
(The details are given in Sec.2.)
In the original Sumino mechanism, the QED correction is cancelled
by $g_F^L g_F^R \log m(A_i^i) \propto -g_F^2 \log m_{ei}$ with
$g_F^L=-g_F^R \equiv g_F$, while in the present 
SUSY model it is done by  $(g_F^{L,R})^2 \log m(A_i^{j})$.  
For this purpose, we will assume the gauge boson masses with an inverted 
hierarchy $m^2(A_i^i)\propto 1/m_{ei}$ as we state in the 
next section.   

In the Sumino model, the gauge boson $A_j^i$ 
couples to  
$$
   (J_\mu^{Sumino})_i^j 
   = \bar{\psi}_L^j \gamma_\mu \psi_{Li}  
   - \bar{\psi}_{Ri} \gamma_\mu \psi_{R}^j ,
   \eqno(1.7)
$$
so that the current-current interactions inevitably cause 
interactions which violate the individual family number $N_F$ 
by $|\Delta N_F|=2$.
This $|\Delta N_F|=2$ effects are somewhat troublesome 
in the low energy phenomenology.
In contrast to the Sumino currents (1.7), our family currents 
are given by the canonical form
$$
(J_\mu)_i^j = \bar{\psi}_L^j \gamma_\mu \psi_{Li}  
+\bar{\psi}_{R}^j \gamma_\mu \psi_{Ri} = \bar{\psi}^j \gamma_\mu \psi_{i},
\eqno(1.8)
$$  
so that the $|\Delta N_F|=2$ effects appear only through a small 
quark family mixing.
 
In summary, the present model has the following characteristics 
compared with the Sumino model: 
(i) Since we can assign the family multiplets as 
$(f_L, f_R) \sim ({\bf 3}, {\bf 3})$ of the U(3),
it is easy to make the model anomaly-less.
(ii) In contrast to the Sumino currents (1.7), we can take
a canonical form of the family currents (1.8).
Therefore, the $|\Delta N_F|=2$ effects appear only through 
a small quark mixing. 
(iii) Family gauge bosons with the lowest and highest 
masses are $A_3^3$ and $A^1_1$, respectively. 
Note that gauge bosons which can couple to the light quarks $u$, $d$ 
and $s$ are only $A^i_j$ with $i,j=1,2$.   
Therefore, we may consider
that the contributions of the family gauge boson exchanges
with $i,j=1,2$ are reduced compared with a conventional 
family gauge boson model. 
This means that we can take a lower value of 
$m(A^3_3)$ (e.g. a few TeV), and may expect to 
observe $A_3^3 \rightarrow \tau^+\tau^-/b\bar{b}/t\bar{t}$ 
via $b\bar{b}/t\bar{t}$ associated production in the LHC. 
In Sec.4, we will investigate a deviation from the $e$-$\mu$ 
universality in the tau lepton decays.
The present data allow the lowest gauge boson mass 
$m(A_3^3)$ to be of the order of a few TeV. 
We will also investigate possible family-number
conserving but lepton flavor violating decays of 
$K$, $D$ and $B$ mesons.
The observations of $K\rightarrow \pi \mu^+ e^-$ and 
$B \rightarrow K \mu^- \tau^+$ are within our reach. 
Thus, we can expect fruitful low energy phenomenology.


\vspace{3mm}
\noindent{\large\bf 2 \ Cancellation mechanism in a SUSY model}

In a SUSY model, the contributions of the family gauge bosons 
in the vertex correction diagram become vanishing, 
so that the original Sumino mechanism does not work.  
On the other hand, those from the wave function renormalization 
diagram still remain:
$$
\delta m_{ei} =  2m_{ei} \sum_j \frac{\gamma_{eij}}{(4\pi)^2}
\log\frac{\mu}{M_{ij}} = m_{ei}
\frac{\alpha_F}{2\pi} \sum_j 
\log\frac{M_{ij}^2}{\mu^2}.
\eqno(2.1)
$$
Here $\gamma_{eij}$ gives the anomalous dimension $\gamma_{ei}$ 
when summed over $j$, 
$$
\gamma_{eij} =- 2 g^2_F \sum_{a} (T^a)_{ij}(T^a)_{ji},\quad
\gamma_{ei} = \sum_j\gamma_{eij} ,  
\eqno(2.2)
$$
where $T^a$ is the generator of the U(3).
Therefore, in the present model, the values of $\varepsilon_i$ defined by 
Eq.(1.3) are given by
$$
\varepsilon_i = \rho \left( \log \frac{m_{ei}^2}{\mu^2} + \zeta \sum_j
\log \frac{M_{ij}^2}{\mu^2} \right) ,
\quad
\rho = \frac{3}{4} \frac{\alpha_{em}}{\pi} , \ \ \  
\zeta=\frac{2}{3} \frac{\alpha_F}{\alpha_{em}} ,
\eqno(2.3)
$$
and $M_{ij}$ are the family gauge boson masses $M_{ij} =m(A_i^j)$
given by
$$
M_{ij}^2  \propto \frac{1}{m_{ei}} +   \frac{1}{m_{ej}} .
\eqno(2.4)
$$
(For a model for $M_{ij}$, see the next section.)

Note that since the gauge bosons $A_i^j$ with $j\neq i$ 
can contribute to the $\varepsilon_i$ term, differently 
from the Sumino model, the QED $\log m_{ei}$ term 
cannot be canceled 
by the gauge boson terms exactly even if we adjust 
the parameter $\zeta$. 
The ratio of $K(m_{ei})$ to $K(m^0_{ei})$ for 
$m_{ei} =m^0_{ei} (1+\varepsilon_0+\varepsilon_i)$
$$
R \equiv  \frac{K(m_{ei})}{K(m^0_{ei})} ,
\eqno(2.5)
$$
is, in general, dependent on the values
$\zeta$ and $\varepsilon_0$. 
(The value of $\varepsilon_0$ is practically not essential
as $|\varepsilon_0| \ll 1$.) 

Next, we investigate the $\zeta$-dependence of the ratio $R$.  
Since the $\varepsilon_0$ term can always be shifted 
by a common value, we shift the $\varepsilon_i$ terms
as $\varepsilon_i \rightarrow \varepsilon_i -\varepsilon_3$
and $\varepsilon_0 \rightarrow \varepsilon_0 +\varepsilon_3$.
Then, we obtain
$$
\frac{1}{\rho} \varepsilon_1 = \log\frac{m_{e1}^2}{m_{e3}^2}
+\zeta \left( \frac{1}{2} \log\frac{m_{e3}^2}{m_{e1}^2}
+\frac{1}{2} \log\frac{m_{e2}^2}{m_{e1}^2}
+\log\frac{1+m_{e1}/m_{e2}}{1+m_{e2}/m_{e3}} \right),
\eqno(2.6)
$$
$$
\frac{1}{\rho} \varepsilon_2 = \log\frac{m_{e2}^2}{m_{e3}^2}
+\zeta \left( \frac{1}{2} \log\frac{m_{e3}^2}{m_{e2}^2}
+\log\frac{1+m_{e1}/m_{e2}}{1+m_{e1}/m_{e3}} \right) ,
\eqno(2.7)
$$
and $\varepsilon_3/\rho =0$.
Here, the first terms in the parentheses in the right hand sides represent 
the contributions of the diagonal gauge fields $A_i^i$, while the succeeding 
terms do those of the off-diagonal ones $A_i^j$ ($j\neq i$). 
As expected, by setting $\zeta$ appropriately ($\zeta=2$), 
the diagonal gauge fields cancel the QED logarithmic terms 
as in the Sumino mechanism, 
but the off-diagonal ones make the cancellation incomplete, 
as $\varepsilon_2/\rho= 2 \log[(1+m_{e1}/m_{e2})/(1+m_{e1}/m_{e3})]
\simeq 2(m_{e1}/m_{e2})$ and $\varepsilon_1/\rho\simeq  
\log(m_{e1}/m_{e2})^2$ in this case. 
Interestingly, however, since $\varepsilon_2$ is quite small and 
safely neglected, the effect on the parameter $K$ which has a mild dependence 
on $m_{e1}$ is relatively suppressed. 
Although the suppression is not sufficient, $R-1={\cal O}(10^{-4})$, 
we expect that the deviation is cancelled by some other effects, 
such as the tau-Yukawa effect, a misarrangement of $\zeta$ for 
instance due to the renormalization group (RG) effects, and so on.

If we want more precise value of $\zeta$ at which $R$ becomes $1$, 
we can obtain it numerically. 
For convenience, we use the following input values: 
the observed charged lepton pole masses \cite{PDG10}
$m^0_{e1}=0.510998910\times 10^{-3}$ GeV, $m^0_{e2}=0.105658367$ GeV,
$m^0_{e3}=1.77682$ GeV; the fine structure constant at $\mu=m_Z$, 
$\alpha(m_Z)=1/127.916$ \cite{PDG10}. 
For the time being, we do not specify an energy scale 
$\Lambda$ at which the formula $K(\Lambda)=2/3$ holds. 
Then, we find that a value of $\zeta$ which gives $R=1$ is
$\zeta=1.752$. 
We can check that the value of $R-1$ at $\zeta=1.752$ is 
insensitive of the value $\varepsilon_0$.
The value $\zeta=1.752$ is near to a value $7/4$.
If we consider $\zeta = 7/4$, we can also show 
that $R-1$ is always smaller than $10^{-5}$ independently of 
the value $\varepsilon_0$. 
Thus, although the present model cannot give rigorous 
cancellation of the $\log m_{ei}$ term, it can practically
give $R=1$ with an accuracy of $10^{-5}$.


\vspace{3mm}

\noindent{\large\bf 3 \ Model} 

A simple way to make a model for the charged leptons 
anomaly-less is to assign the lepton doublet $\ell$ 
and charged lepton singlet $e^c$
to ${\bf 3}$ and ${\bf 3}^*$ of the U(3) family symmetry, respectively, 
in contrast to 
the Sumino model \cite{Sumino09PLB, Sumino09JHEP} where 
both have been assigned to ${\bf 3}$.  
If we adopt a yukawaon model\cite{yukawaon09},
the would-be Yukawa interaction
for the charged lepton sector 
is given by $(y_e/\Lambda) \ell_i (Y_e)^i_j e^{cj} H_d$.
Then, however, another term
$(y'_e/\Lambda) \ell_i (Y_e)^j_j e^{ci} H_d$
is also allowed by the symmetry, and there are no reasons to forbid it.
(This problem always appear when we take a model with 
$\ell \sim {\bf 3}$ and $e^c \sim {\bf 3}^*$.)
Therefore, in the present paper, we do not adopt such a 
yukawaon model. 
Following to the Sumino model, we will consider 
a bilinear contribution 
$(\bar{\Phi}_e)^i_\alpha (\Phi_e)^{\alpha}_j$
instead of $(Y_e)^i_j$. 
To be more concrete, we assign 
$(\bar{\Phi}_e)^i_\alpha \sim ({\bf 3}^*, {\bf 3})$ 
and $(\Phi_e)^{\alpha}_j \sim ({\bf 3}, {\bf 3}^*) $ of
U(3)$\times$U(3)$'$ family symmetries. 
Note that we consider U(3)$'$ instead of O(3) 
in the Sumino model.
This allows us to take a flavor basis in which 
$\langle \Phi_e \rangle$ is diagonal.
Hereafter, we simply denote $\Phi_e$ and $\bar{\Phi}_e$ 
in the present model as $\Phi$ and $\bar{\Phi}$, 
respectively. 

We take the following would-be Yukawa interaction terms
$$   
W_Y = y_\ell \ell_i \bar{\Phi}^i_\alpha \bar{L}^\alpha
+ y_{Hd} {L}_\alpha H_d {E}^{c\alpha} + y_e \bar{E}^c_\alpha 
\Phi^\alpha_j e^{cj}
+ M_E E^{c\alpha}\bar{E}^c_\alpha 
+M_L L_\alpha  \bar{L}^\alpha ,
\eqno(3.1)
$$
where $\ell_i =(\nu_i, e_i)$, $e^{ci}$, 
$L_\alpha = (N_\alpha, E_\alpha)$, $E^{c\alpha}$, 
$\bar{L}^\alpha = (\bar{N}^\alpha, \bar{E}^\alpha)$ and 
$\bar{E}^c_\alpha$, have the electric charges $(0,-1)$, $+1$, 
$(0,-1)$, $+1$, $(0,+1)$ and $-1$, respectively. 
Here, $(E^c, \bar{E}^c)$ and $(L,\bar{L})$ are vector-like 
SU(2)$_L$ singlets and doublets, respectively.  
Then, we obtain the following effective superpotential
$$
W^{eff}_Y = \frac{y_{Hd} y_\ell y_e}{M_E M_L}
 \ell_i \bar{\Phi}^i_\alpha 
 \Phi^\alpha_j {e}^{c\, j} H_d .
\eqno(3.2)
$$
Note that the counterpart of the $y'$-term 
${\rm Tr}[\bar{\Phi} \Phi] \ell_i {e}^{c\, i} H_d$
is not generated at the tree level, 
and then, in great contrast to non-SUSY models, is protected against 
the radiative corrections thanks to the nonrenormalization theorem. 
To be complete, we should forbid a term $\ell_i {e}^{c\, i} H_d$ 
and hopefully the above effective nonrenormalizable term generated 
already at the cutoff scale $\Lambda$. 
The former term can be forbidden by a U(1)$_S$ symmetry as usual. 
A concrete example of assignment of the charge and other quantum numbers 
(including those of the U(1)$_R$ symmetry that the superpotential 
considered here has) 
are shown in Table 1.
The latter term may be forbidden effectively by assuming that the cutoff scale is 
large enough, or 
by replacing for instance the mass $M_E$ with a VEV of a field $S$ charged 
under the U(1)$_S$ symmetry. 
Here, we employ the former way for simplicity and consider only renormalizable 
terms, while we also introduce the field $S$ for the later convenience.

To avoid the massless Nambu-Goldstone mode, 
we assume that the U(1)$_S$ charge conservation is broken by a soft term
$$
W_{br} = \mu_S S \theta_S - \varepsilon \mu_S^2 \theta_S ,
\eqno(3.3)
$$
where $S$ and $\theta_S$ are family singlets, 
and $\varepsilon$ has been put in order to denote that 
the $R$ charge conservation in the term (3.3) is softly 
broken with a small extent $\varepsilon$.
The superpotential (3.3) leads to
$ \langle S \rangle = \varepsilon \mu_S$.

The effective superpotential (3.2) reduces to the charged lepton Yukawa 
interaction when $\Phi$ and $\bar\Phi$ acquire VEVs. 
The magnitude of $\langle \Phi \rangle \langle \bar{\Phi} \rangle$   
is set by 
$$
W_\Phi = \lambda_1 \Phi^\alpha_i \bar{\Phi}^i_\alpha \theta_\Phi
-\lambda_2 S^2 \theta_\Phi ,
\eqno(3.4)
$$
where $\theta_\Phi$ is a family singlet field, 
which leads to
$$
{\rm Tr}[ \langle \Phi \rangle \langle \bar{\Phi} \rangle]
= \varepsilon^2  \frac{\lambda_2}{\lambda_1} \mu_S^2.
\eqno(3.5)
$$ 
As mentioned, we do not discuss how appropriate forms (1.5) 
of the VEVs, $\langle \Phi \rangle$ and $\langle \bar{\Phi} \rangle$, 
are obtained 
but just assume that a superpotential $W_K(\Phi,\bar\Phi,\cdots)$ leads them.
Here, in addition, we assume that $W_K$ is invariant\footnote{
The superpotential (3.1) does not possess this invariance.
}
under the exchange of the U(3) and the U(3)$'$, which exchanges $\Phi$ 
and $\bar\Phi$, and that the VEVs respect this $S_2(=Z_2)$ symmetry:
$$
\langle {\Phi} \rangle = \langle \bar{\Phi} \rangle ={v}_{\Phi} {Z} 
= v_\Phi \, {\rm diag}(z_1, z_2, z_3).
\eqno(3.6)
$$
Here, in the last equality, we have used the U(3) and U(3)$'$ degrees of freedom 
to diagonalize $\langle {\Phi} \rangle$, and
the parameters $z_i$ are normalized as $z_1^2+z_2^2+z_3^2=1$,
without loosing generality.
With this notation, from Eq.(3.5), we obtain
$$
v_\Phi^2 = \varepsilon^2 \frac{\lambda_2}{\lambda_1} \mu_S^2. 
\eqno(3.7)
$$

Next, we investigate a superpotential for the field $\Psi$
whose VEV $\langle \Psi \rangle$ gives an inverted mass 
hierarchy (2.4): 
$$
W_{\Phi\Psi} = \left( \lambda_A \bar{\Psi}^i_\alpha {\Phi}^\alpha_j
+ \bar{\lambda}_A \bar{\Phi}^i_\alpha {\Psi}^\alpha_j \right) 
(\Theta_A)^j_i 
+ \left(\lambda'_A \bar{\Psi}^i_\alpha \Phi^\alpha_i 
+ \bar{\lambda}'_A \bar{\Phi}^i_\alpha {\Psi}^\alpha_i
-\mu_A S \right) (\Theta_A)^j_j 
$$
$$
+ \left( \lambda_B {\Phi}^\alpha_i \bar{\Psi}^i_\beta 
+ \bar{\lambda}_B {\Psi}^\alpha_i \bar{\Phi}^i_\beta \right) 
(\Theta_B)^\beta_\alpha  
+ \left( \lambda'_B {\Phi}^\alpha_i \bar{\Psi}^i_\alpha
+ \bar{\lambda}'_B {\Psi}^\alpha_i \bar{\Phi}^i_\alpha
-\mu_B S \right) (\Theta_B)_\beta^\beta .
\eqno(3.8)
$$
Again, we impose that $W_{\Phi\Psi}$ is $S_2$ invariant, i.e. 
$\lambda_A = \bar{\lambda}_B$, $\bar{\lambda}_A =\lambda_B$, 
$\lambda'_A = \bar{\lambda}'_B$, $\bar{\lambda}'_A =\lambda'_B$ 
and $\mu_A = \mu_B$.
Then, from the $F$-flatness conditions, we obtain 
$$
\langle {\Psi} \rangle = \langle \bar{\Psi} \rangle = v_{\Psi} {Z}^{-1}, 
\eqno(3.9)
$$
which satisfies the $D$-term condition too. 
We also obtain
$$
v_\Phi {v}_\Psi = 3 \varepsilon \frac{ \mu_S \mu_A}{\lambda_A+3\lambda'_A
+ \bar{\lambda}_A+3\bar{\lambda}'_A} .
\eqno(3.10)
$$
Comparing with Eq.(3.7), we see
$$
\frac{{v}_\Phi}{{v}_\Psi} = \varepsilon k
\sim O(\varepsilon) ,
\eqno(3.11)
$$
where $k=(1/3) (\lambda_2/\lambda_1) (\lambda_A + 3\lambda'_A 
+\bar{\lambda}_A+3\bar{\lambda}'_A)(\mu_S/\mu_A)$.
Since the charged lepton masses $m_{ei}$ are
given by
$$
M_e \equiv {\rm diag}(m_{e1}, m_{e2}, m_{e3})
\propto v_\Phi^2 Z^2  
= v_\Phi^2 {\rm diag}(z_1^2, z_2^2, z_3^2)  ,
\eqno(3.12)
$$
from Eq.(3.2), 
the parameter values of $z_i$ are given by 
$$
z_i =\frac{\sqrt{m_{ei}}}{\sqrt{ m_{e1}+m_{e2}+m_{e3}}} ,
\eqno(3.13)
$$
where $(m_{e1}, m_{e2}, m_{e3})=(m_e, m_\mu, m_\tau)$. 
The explicit values of $z_i$ are given by
$$
(z_1,z_2,z_3) = (0.016473, 0.23688, 0.97140) .
\eqno(3.14)
$$

Thus, we can approximately estimate the family gauge boson masses 
$m(A^i_j)$ as follows
$$
M_{ij}^2 \equiv m^2(A^i_j) = \frac{1}{2} g_F^2 \left[ \sum_\alpha 
(\langle (\Psi^\dagger)^i_\alpha \rangle
\langle {\Psi}^\alpha_i \rangle + 
\langle {\bar{\Psi}}^i_\alpha \rangle
\langle (\bar{\Psi}^\dagger)^\alpha_i \rangle ) 
+(i\rightarrow j) 
+  O(\varepsilon^2)  \right]
$$
$$
\simeq 
 g_F^2 v_\Psi^2 \left( 
\frac{1}{z_i^2} + \frac{1}{z_j^2} \right)
\propto 
 \left(
\frac{1}{m_{ei}} + \frac{1}{m_{ej}} \right),
\eqno(3.15)
$$
if the mixing between the U(3) gauge boson and the U(3)$'$ one can be neglected. 
This happens when the latter gauge boson is sufficiently heavy and we assume such 
a case. 
Namely, we assume another sector that breaks U(3)$'$ at a high scale which is 
basically decoupled from the sector we have discussed above. 

In this case, in addition to the interactions in the superpotential (3.1), 
the gauge interaction (below the U(3)$'$ breaking scale) violates the $S_2$ 
symmetry assumed in $W_K$ and $W_{\Phi\Psi}$, and the RG effects modify the 
$S_2$ relations shown above Eq.(3.9). 
Nevertheless, amazingly, the nonrenormalization theorem protects the VEV relations 
(3.6) and (3.9) against the RG effects, which justifies the above discussion.

\begin{table}

\begin{tabular}{c|cccccccccccccccc} \hline
     & $\ell$ & $e^c$ & $H_d$ & $L$ & $\bar{L}$ & ${E}^c$ & $\bar{E}^c$ 
& $\Phi$ & $\bar{\Phi}$ & $\Psi$ & $\bar{\Psi}$ & $\theta_\Phi$
& $\Theta_A$ & $\Theta_B$ & $S$ & $\theta_S$ \\ \hline   
SU(2)$_L$ & ${\bf 2}$ & ${\bf 1}$ & ${\bf 2}$ 
& ${\bf 2}$ & ${\bf 2}$ & ${\bf 1}$ & ${\bf 1}$ & 
 ${\bf 1}$ & ${\bf 1}$ & ${\bf 1}$ & ${\bf 1}$ & ${\bf 1}$ 
& ${\bf 1}$ & ${\bf 1}$ & ${\bf 1}$ & ${\bf 1}$ \\  
U(3) & ${\bf 3}$ & ${\bf 3}^*$ & ${\bf 1}$ & ${\bf 1}$
& ${\bf 1}$ & ${\bf 1}$ & ${\bf 1}$ & 
${\bf 3}$ & ${\bf 3}^*$ & ${\bf 3}$ & ${\bf 3}^*$ & ${\bf 1}$
& ${\bf 8}+{\bf 1}$ & ${\bf 1}$ & ${\bf 1}$ & ${\bf 1}$ \\ 
U(3)$'$ & ${\bf 1}$ & ${\bf 1}$ & ${\bf 1}$ 
& ${\bf 3}$ & ${\bf 3}^*$ & ${\bf 3}^*$ & ${\bf 3}$ &
${\bf 3}^*$ & ${\bf 3}$ & ${\bf 3}^*$ & ${\bf 3}$ & ${\bf 1}$
& ${\bf 1}$ & ${\bf 8}+{\bf 1}$  & ${\bf 1}$ & ${\bf 1}$ \\
U(1)$_S$ & $0$  & $0$  & $-1$ & $1$ & $-1$ & $0$ & $-1$ &
$1$ & $1$ & $0$ & $0$  
& $-2$ & $-1$ & $-1$ & $1$ &  $-1$ \\ 
U(1)$_R$ & $1$ & $1$ & $0$  & $1$ & $1$  & $1$ & $1$ & 
$0$ & $0$  & $0$ & $0$ & $2$ & $2$  & $2$ & $0$ & $2$ \\
\hline
\end{tabular}

\caption{The fields in the present model and their quantum numbers.}

\end{table}

So far, we have not discussed the neutrino mass matrix, because 
the purpose of the present paper is to discuss how to apply the 
Sumino mechanism to a SUSY model. 
Here, we would like to give a brief comment on the neutrino mass 
matrix.
In order to obtain neutrino masses, we add, for instance, a new field  
${N}^c$, which is SU(2)$_L$ doublet with ${\bf 3}^*$ of the family
symmetry SU(3)$'$. 
(However, we do not consider the vector-like partner $\bar{N}^c$ 
unlike the case $(E^c, \bar{E}^c)$.
Therefore, in order to make the model anomaly free, 
some additional fields are needed. 
In this paper, we do not comment on it.)  
The field generates the following superpotential terms 
in addition to Eq.(3.1):
$$
y_{Hu} L_\alpha H_u {N}^{c\alpha} + y_M {N}^{c\alpha} 
(Y_M)_{\alpha\beta} {N}^{c\beta} .
\eqno(3.16)
$$
The superpotential terms (3.1) and (3.16) leads to the effective 
neutrino mass matrix as Eq.(3.2):
$$
W_\nu^{eff} = \frac{y_{Hu} y_\ell}{y_M M_L^2} (\ell_iH_u) \bar{\Phi}_\alpha^i
(\langle Y_M \rangle^{-1})^{\alpha\beta} \bar{\Phi}_\beta^j (\ell_jH_u).
\eqno(3.17)
$$
Here, we consider that the VEV value $\langle Y_M \rangle_{\alpha\beta}$
breaks SU(3)$'$ symmetry at a higher scale $\Lambda'$ 
($\Lambda' \gg \Lambda$).
This realizes the above assumption used in Eq.(3.15), and 
is a reason that we will not discuss SU(3)$'$ family symmetry 
gauge boson effects in the following low energy scale phenomenology.

\vspace{3mm}

\noindent{\large\bf 4 \ Possible effects of the family gauge bosons}

Since the gauge boson masses are given by Eq.(3.15), 
we obtain the following hierarchical structure: 
$$
\begin{array}{l}
2 g_F^2 v_\Psi ^2 \frac{1}{z_1^2} = M^2_{11}  
\simeq 2 M^2_{12} \simeq 2 M^2_{13} , \\
2 g_F^2 v_\Psi ^2 \frac{1}{z_2^2} = M^2_{22} 
\simeq 2 M^2_{23} , \\
2 g_F^2 v_\Psi ^2 \frac{1}{z_3^2} = M^2_{33} . \\
\end{array}
\eqno(4.1)
$$

The family currents in the Sumino model are given by Eq.(1.7), 
while, in the present model, those are given by 
$$
(J_\mu)_j^i = \bar{e}_L^i \gamma_\mu e_{L j}+ 
\bar{e}_R^i \gamma_\mu e_{R j}+ \cdots =  
\bar{e}^i \gamma_\mu e_j+ \cdots,
\eqno(4.2)
$$
where, for convenience, we have denoted only charged lepton sector explicitly.
Note that the effective current-current interactions   
in the Sumino model induce $\Delta N_F =2$ interactions, 
while those in the present model do not induce such  $\Delta N_F =2$ 
interactions.  
In the present model, however, since the family number is defined in a diagonal basis
of the charged lepton mass matrix, 
in general, quark mixings 
appear, so that family number violating modes will be observed   
through $U_u \neq {\bf 1}$ and $U_d \neq {\bf 1}$,
where $U_q$ ($q=u,d$) is defined by $U_{Lq}^\dagger M_q U_{Rq} = D_q$ 
($D_q$ is a diagonal matrix).   
For example, family currents in the down-quark sector are given by
$$
(J_\mu^{(d)})_j^i = (\bar{d}^{0 i}_L \gamma_\mu d^0_{Lj}) + 
(\bar{d}^{0 i}_R \gamma_\mu d^0_{Rj} )
=(U_{Ld}^\dagger )^i_k (U_{Ld})^l_j (\bar{d}_L^k \gamma_\mu d_{L l} )
+(L \rightarrow R) ,
\eqno(4.3)
$$ 
so that the effective Hamiltonians for semileptonic modes and 
nonleptonic modes, $H_{SL}$ and $H_{NL}$, are given by
$$
H_{SL}^{eff} = \sum_{i,j,k,l} \frac{G_{ij}}{\sqrt{2}} 
(U_d^\dagger)^i_k (U_d)^l_j
(\bar{d}^k \gamma_\mu d_l)(\bar{e}^j \gamma^\mu e_i) ,
\eqno(4.4)
$$
$$
H_{NL}^{eff} = \sum_{i,j,k,l, m,n} \frac{G_{ij}}{\sqrt{2}}
(U_d^\dagger)^i_k (U_d)^l_j (U_d^\dagger)^j_m (U_d)^n_i 
(\bar{d}^k \gamma_\mu d_l)(\bar{d}^m \gamma^\mu d_n) ,
\eqno(4.5)
$$
respectively, where 
$G_{ij}/\sqrt{2} = g_F^2/2 M_{ij}^2 \simeq z_j^2/2 v_\Psi^2$,
and, for simplicity, we have put $U_{Ld}=U_{Rd}\equiv U_d$.   
In this section, we investigate possible phenomenology of 
the flavor violating modes, and discuss the scale of the gauge bosons.

Usually, the most strict constraint comes from the observed 
$K^0$-$\bar{K}^0$ mixing. 
However, this $\Delta N_F =2$ transition occurs only via the 
down-quark mixing $U_d \neq {\bf 1}$, so that the constraint is
highly dependent on the quark mass matrix model. 
In this paper, we do not discuss a model about the quark mixing.
Only modes that are independent of the quark mixing structures are
pure leptonic decays $e_i \rightarrow e_j + \bar{\nu}_j + \nu_i$.
Therefore, first, let us investigate these pure leptonic decays 
based on the present model. 
The effective interactions via the family gauge boson exchanges are
given by
$$
\frac{G_{ij}}{\sqrt{2}} (\bar{\nu}_i \gamma_\mu \nu_j)
(\bar{e}_j \gamma^\mu e_i) , 
\eqno(4.6)
$$
against the conventional weak interactions
$$
\frac{G_F}{\sqrt{2}} (\bar{e}_j \gamma_\mu (1-\gamma_5) \nu_j)
(\bar{\nu}_i \gamma^\mu (1-\gamma_5) e_i) , 
\eqno(4.7)
$$
where $G_F/\sqrt{2} = g_W^2/8 M_W^2 = 1/2 v_W^2$ 
($v_W = 246$ GeV).
By using the Fierz transformation, we obtain effective coupling 
constants (for the definitions, see \cite{decay_param}) 
in the current-current interactions 
$$
g^V_{LL} = 1+\epsilon_j , \ \ \ g^V_{RR} = 0 ,
\ \ \  g^S_{LR} = -2 \epsilon_j , \ \ \ g^S_{RL} = 0 ,
\eqno(4.8)
$$
where $\epsilon_j \simeq (1/4) z_j^2 (v_W/v_\Psi)^2$, 
and we have considered a case that the observed neutrinos 
are Majorana types. 
The result (4.8) gives the decay parameters
 \cite{decay_param} $\eta=0$, $\rho = 3/4$, $\delta =3/4$ and
 $\xi \simeq  1 -2 \epsilon_j^2$.
Regrettably, the results for $\eta$, $\rho$ and $\delta$ are
identical with those in the standard model (SM) and the deviation
of $\xi$ from $\xi^{SM} =1$ is too small to observe. 
On the other hand, in relation to the branching ratios, we predict
$$
R_\tau \equiv \frac{1+\epsilon_\mu}{1+\epsilon_e }
=\left[ 
\frac{B(\tau^- \rightarrow \mu^- \bar{\nu}_\mu \nu_\tau)}{
B(\tau^- \rightarrow e^- \bar{\nu}_e \nu_\tau)}
\frac{f(m_e/m_\tau)}{f(m_\mu/m_\tau)} \right]^{1/2} ,
\eqno(4.9)
$$
where $f(x)$ has been defined by 
$f(x)=1-8 x^2 +8 x^6-x^8 -12 x^4 \ln x^2$ and 
${f(m_e/m_\tau)}/{f(m_\mu/m_\tau)}= 1.028215$.
Since $\epsilon_e \simeq z_1^2 r^2/4 = 6.8 
\times 10^{-5} r^2$
and $\epsilon_\mu \simeq z_2^2 r^2/4 = 1.4  
\times 10^{-2} r^2$ [$r \equiv v_W/v_\Psi$], we expect 
a deviation 
$\Delta R_\tau \equiv R_\tau -1 \simeq \epsilon_\mu$. 
Present experimental values \cite{PDG10} 
$B(\tau^- \rightarrow \mu^- \bar{\nu}_\mu \nu_\tau )=
(17.39 \pm 0.04) \%$ and 
$B(\tau^- \rightarrow e^- \bar{\nu}_e \nu_\tau )=
(17.82 \pm 0.04) \%$ give 
$$
R_\tau^{exp} = 1.0017 \pm 0.0016 ,
\eqno(4.10)
$$ 
i.e. $\epsilon_\mu \simeq 0.0017 \pm 0.0016$.
This result seems to be in favor of the inverted 
gauge boson mass hierarchy although it is just at 1 $\sigma$ level.
(If the gauge boson masses take a normal hierarchy,
$R_\tau$ will show $R_\tau < 1$.)   
However, if we take the central value $\Delta R_\tau \sim 0.0017$, 
it means $r \sim 0.35$. 
This value corresponds to $v_\Psi \sim 0.7$ TeV which is ruled out 
by Kaon rare decays as we will see next. 
At present, we should not take the value (4.10) rigidly.  
If we speculate  $r \sim 10^{-1}$ ($v_\Psi$ of a few TeV ),
we may expect a sizable deviation 
$\Delta R_\tau \simeq \epsilon_\mu \simeq  z_2^2 r^2/4 \sim 10^{-4}$ 
from the $e$-$\mu$ universality in the tau lepton decays.   
We expect that the observation $\Delta R_\tau \simeq 10^{-4}$ 
will be accomplished by a tau-factory in the near future.

Next, we direct our attention to family
number conserved modes in the limit of no quark mixing.
Predicted values for those modes are insensitive to 
the explicit values of $U_d$ and $U_u$ 
as far as they are not so large. 
In particular, we investigate rare decays of pseudo-scalar mesons 
$0^- \rightarrow 0^- + e_i +\bar{e}_j$ ($i\neq j$) 
with $\Delta N_F =0$,
e.g. $K^+ \rightarrow \pi^+ \mu^+ e^-$, 
$B^+ \rightarrow K^+ \tau^+ \mu^-$, and so on.  
(Since our currents are pure vectors, they cannot
contribute to decays $0^-  \rightarrow e_i +\bar{e}_j$.)
When we neglect the $CP$ violation effects and the electromagnetic
mass difference of pseudo-scalar mesons, we can predict 
the following branching ratios:
$$
B(K^+\rightarrow \pi^+ \mu^+ e^-) \simeq 2 z_1^4 r^4
\frac{1}{2|V_{us}|^2} B(K^+\rightarrow \pi^0 \mu^+ \nu_\mu)
=4.88 \times 10^{-8} r^4 ,
\eqno(4.11)
$$
$$
B(K^0 \rightarrow \pi^0 \mu^+ e^-) \simeq \frac{1}{2} z_1^4 r^4
\frac{1}{2|V_{us}|^2} B(K^0 \rightarrow \pi^- \mu^+ \nu_\mu)
=9.82 \times 10^{-8} r^4 ,
\eqno(4.12)
$$
$$
B(D^+\rightarrow \pi^+ \mu^- e^+) \simeq  z_1^4 r^4
\frac{1}{2|V_{cs}|^2} B(D^+\rightarrow \bar{K}^0 \mu^+ \nu_\mu)
\frac{f(m_\pi/m_D)}{f(m_K/m_D)}
=5.83 \times 10^{-9} r^4 ,
\eqno(4.13)
$$
$$
B(D^0 \rightarrow \pi^0 \mu^- e^+) \simeq \frac{1}{2} z_1^4 r^4
\frac{1}{2|V_{cs}|^2} B(D^0 \rightarrow {K}^- \mu^+ \nu_\mu)
\frac{f(m_\pi/m_D)}{f(m_K/m_D)}
=1.03 \times 10^{-9} r^4 ,
\eqno(4.14)
$$
$$
B(B^+\rightarrow K^+ \mu^- \tau^+) \simeq  z_2^4 r^4
\frac{1}{2|V_{cb}|^2} B(B^+\rightarrow \bar{D}^0 \tau^+ \nu_\tau)
\frac{f(m_K/m_B)}{f(m_D/m_B)}
=1.51 \times 10^{-2} r^4 ,
\eqno(4.15)
$$
$$
B(B^0\rightarrow K^0 \mu^- \tau^+) \simeq  z_2^4 r^4
\frac{1}{2|V_{cb}|^2} B(B^0\rightarrow {D}^- \tau^+ \nu_\tau)
\frac{f(m_K/m_B)}{f(m_D/m_B)}
=2.37 \times 10^{-2} r^4 ,
\eqno(4.16)
$$
where $r$ and $f(x)$ have been defined below Eq.(4.9).   
(For simplicity, we have used approximate relation in the limit
of massless charged leptons. Therefore, the numerical results 
should not be taken rigidly.) 
In Eq.(4.12), under the approximation of neglecting $CP$ violation, 
we read $B(K^0 \rightarrow \pi^- \mu^+ \nu_\mu)$ as 
$B(K_L \rightarrow \pi^\pm \mu^\mp \nu_\mu)=
(1/2) B(K^0 \rightarrow \pi^- \mu^+ \nu_\mu) 
+(1/2) B(\bar{K}^0 \rightarrow \pi^+ \mu^- \bar{\nu}_\mu)
= B(K^0 \rightarrow \pi^- \mu^+ \nu_\mu) 
= B(\bar{K}^0 \rightarrow \pi^+ \mu^- \bar{\nu}_\mu)$
(and also $B(K^0 \rightarrow \pi^0 \mu^+ e^-)$ as
$B(K_L \rightarrow \pi^0 \mu^\pm e^\mp)$). 
In the numerical results in Eqs.(4.11) - (4.16), 
we have used the observed values \cite{PDG10} $B(K^+\rightarrow 
\pi^0 \mu^+ \nu_\mu) =3.353 \times 10^{-2}$, 
$B(K_L \rightarrow \pi^\mp \mu^\pm \nu_\mu)=0.2704$, 
$B(D^+\rightarrow \bar{K}^0 \mu^+ \nu_\mu)= 9.4 \times 10^{-2}$,
$B(D^0 \rightarrow {K}^- \mu^+ \nu_\mu) =3.31 \times 10^{-2}$,
$B(B^+\rightarrow \bar{D}^0 \tau^+ \nu_\tau)=7 \times 10^{-3}$ 
and $B(B^0\rightarrow {D}^- \tau^+ \nu_\tau)=1.1 \times 10^{-2}$.
For reference, we list the predicted values for $v_W/v_\Psi \sim 10^{-1}$ 
(and present experimental upper limits \cite{PDG10})  
as follows:
$$
\begin{array}{lll}
B(K^+\rightarrow \pi^+ \mu^+ e^-) & \sim 5 \times 10^{-12} & 
( < 1.3 \times 10^{-11}) , \\
B(K_L \rightarrow \pi^0 \mu^\pm e^\mp) & \sim 1 \times 10^{-11} & 
( <7.6 \times 10^{-11} ) , \\
B(D^+\rightarrow \pi^+ \mu^- e^+) & \sim 6 \times 10^{-13} & 
( <3.4 \times 10^{-5} ) , \\ 
B(D^0 \rightarrow \pi^0 \mu^- e^+) & \sim 1 \times 10^{-13} &
( <8.6 \times 10^{-5}) , \\ 
B(B^+\rightarrow K^+ \mu^- \tau^+) & \sim 2 \times 10^{-6} & 
( <7.7 \times 10^{-5} ) , \\ 
B(B^0\rightarrow K^0 \mu^- \tau^+) & \sim 2 \times 10^{-6} & 
( {\rm no\ data} ) . 
\end{array}
\eqno(4.17)
$$
We also predict $B(K_L \rightarrow \pi^0 \nu_e \bar{\nu}_\mu) 
\simeq B(K_L \rightarrow \pi^0 \mu e)/2$. 
We show the predicted branching ratios $B(P\rightarrow P'
e_i \bar{e}_j)$ versus $v_F \equiv v_\Psi$ in Fig.1.
Therefore, if $v_\Psi$ is a few TeV, observations 
of the lepton-flavor violating $K$- and $B$-decays with 
$\Delta N_F =0$ will be within our reach. 


\begin{figure}
{\scalebox{0.4}{\includegraphics{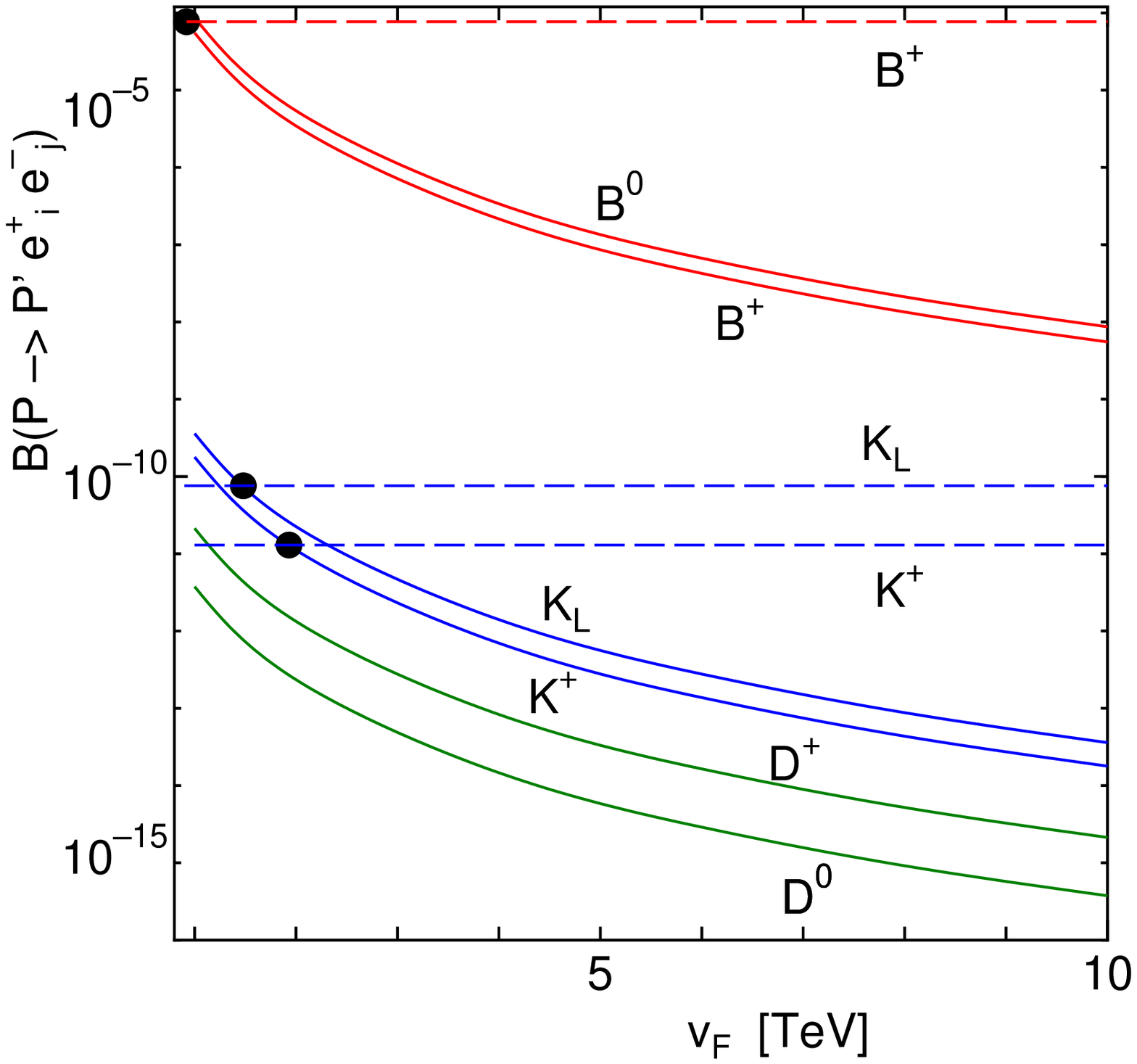}} }

\begin{quotation}\small
{\bf Fig.~1}  Predicted branching ratios $B(P \rightarrow P' e^+_i e^-_j)$
versus the VEV value $v_F \equiv v_{\Psi}$.
The marks $\bullet$ and the dashed lines denote present lower limits 
of the observed branching ratios. 
\end{quotation}
\end{figure}

We may also expect to observe the lightest family 
gauge boson $A_3^3$ if its mass is 
a few TeV.
For simplicity, we neglect the up- and down-quark 
mixings, i.e. $(u_1, u_2, u_3) \simeq (u, c, t)$ and 
$(d_1, d_2, d_3) \simeq (d, s, b)$. 
The observation is practically the same as that 
for $Z'$ boson (for a review, see Ref.\cite{Z-prime}).
Although in the conventional $Z'$ model, $Z'$ couples 
to the fermions of all flavors, while the $A_3^3$ boson
couples only to $\tau^+ \tau^-$, $\nu_\tau \bar{\nu_\tau}$,
$b \bar{b}$ and $t \bar{t}$, so that the branching ratios
are given by
$$
B(A_3^3 \rightarrow \tau^+ \tau^-) \sim 
2 B(A_3^3 \rightarrow \nu_\tau \bar{\nu_\tau}) \sim 
\frac{1}{3} B(A_3^3 \rightarrow b \bar{b}) \sim 
\frac{1}{3} B(A_3^3 \rightarrow t \bar{t}) \sim \frac{2}{15}.
\eqno(4.18)
$$
We may expect to observe a peak of $\tau^+ \tau^-$ 
(but no peak in $e^+ e^-$ and $\mu^+ \mu^-$) in  
$pp \rightarrow gg X \rightarrow  A_3^3X \rightarrow \tau^+ \tau^-X$ 
at the LHC and $e^+ e^- \rightarrow Z^*/\gamma^* 
\rightarrow  A_3^3X \rightarrow \tau^+ \tau^-X$ 
at the ILC, although these cross sections of the $A_3^3$ 
productions are small compared with that of the $Z'$ production. 
Similar discussion can be applied to hadronic jets instead 
of $\tau^+ \tau^-$. 

Finally, we would like to comment on a constraint from 
the observed $K^0$-$\bar{K}^0$ mixing. 
As we stated previously, contributions from exchanges of 
the U(3) family gauge bosons to the $K^0$-$\bar{K}^0$ mixing 
depend on the magnitudes of the family mixing $U_d$ in 
the down-quark sector.  
At present, we know the observed values of the CKM mixing 
$V_{CKM} = U^\dagger_u U_d$,
while we do not know the value of $U_d$. 
Tentatively, let us assume  
that the CKM mixing is dominantly given by the down-quark
mixing, i.e. $U_d \simeq V_{CKM}$.
Then, the dominant contribution comes from
the exchange of the family gauge boson $A_2^2$: 
$({g_F^2}/{2 M^2_{22} }) (V_{us}^* V_{cs})^2  =
({z_2^2}/{4 v_\Psi^2})  (V_{us}^* V_{cs})^2 
= (v_\Psi^2)^{-1}  \times 6.76 \times 10^{-4}$.
In the present model, the $CP$ violating effect in the dominant contribution 
is negligibly small. 
Since the standard model gives $\Delta m_K^{SM} \simeq (7/6 - 5/6)
\Delta m_K^{exp}$\cite{Inami-Lim} 
($\Delta m_K^{exp} = (3.483 \pm 0.006) \times 10^{-12}$
 MeV \cite{PDG10}), we consider that a contribution from 
new physics \cite{Bai} is $|\Delta m_K^{NewPhys}| < \Delta m_K^{exp}/6$. 
If we assume the vacuum-insertion approximation, we obtain
a constraint $v_\Psi \gtrsim {\cal O}(10^{5})$ GeV,
which suggests that the lightest gauge boson mass should also satisfy 
$m(A_3^3) \gtrsim {\cal O}(10^{5})$ GeV. 
This result contradicts our speculation $r \sim 10^{-1}$.   
If this speculation is confirmed in future observations, 
we must build a quark mass matrix model with 
$U_d \simeq {\bf 1}$ in the diagonal basis of the charged 
lepton mass matrix $M_e$, especially with 
$(U_d)_{12}\simeq 0$. 
(This means that the down-quark mass matrix $M_d$ takes 
a similar structure except for a unit matrix term, i.e.
$M_d \simeq k_d M_e + m_0 {\bf 1}$.)



\vspace{3mm}

\noindent{\large\bf 5 \ Concluding remarks} 

In conclusion, we have proposed a family gauge boson model 
with inverted hierarchical masses.
The model has been motivated to give a SUSY scenario of 
the Sumino's cancellation mechanism between the radiative
correction due to the photon and that due to the family gauge bosons.
As stated in the end of Sec.1, the present model has many 
characteristics compared with the Sumino model:
(i) It is easy to build a model that is anomaly free, 
since the model takes the canonical assignments of the U(3) 
family.
(ii) The dangerous $\Delta N_F=2$ interactions do not appear 
in the limit of no quark mixing.
(iii)  The family gauge bosons with the inverted hierarchical 
masses offer a new view for the low energy phenomenology.    
(iv) Since our model is based on a SUSY scenario, 
the VEV relations are kept (up to the SUSY breaking effects) although 
in this paper we did not 
discuss the derivations of the relation (1.1) and so on. 

If we take the mass relation (1.1) seriously and we want
to apply the Sumino mechanism to a SUSY scenario, 
the present model will be a promising model as an 
alternative one of the Sumino model.
Whether the gauge boson mass hierarchy is inverted or normal will be 
confirmed by observing the direction of the deviation form the $e$-$\mu$ 
universality in the pure leptonic tau decays.
The present experimental result, $R_\tau =1.0017 \pm 0.0016$,
is in favor of the inverted mass hierarchy although the
error is still large.
Since we speculate that the lightest gauge boson mass is a few TeV,
we expect the deviation $\Delta R_\tau = R_\tau -1 \sim 10^{-4}$.
A tau factory in the near future will confirm this deviation.
In addition, some lepton flavor violating $K$- and $B$-decays, 
e.g. $K^+ \rightarrow \pi^+ \mu^+ e^-$ and 
$B^+ \rightarrow K^+ \tau^+ \mu^-$, will be within our reach. 
We also expect a direct observation of $\tau^+\tau^-$/$b\bar{b}$/$t\bar{t}$
decay modes while no excesses in the $e^+e^-$/$\mu^+\mu^-$ modes 
in the LHC and the ILC.

 \vspace{3mm}

{\large\bf Acknowledgments}

The authors would like to thank Y.~Sumino for valuable 
and helpful conversations, and also M.~Tanaka and K.~Tsumura for 
helpful discussions for lepton flavor violating and collider 
phenomenology.
One of the authors (YK) is supported by JSPS (No.\ 21540266).


\end{document}